\documentclass[pra,floatfix,amsmath,twocolumn]{revtex4}
\usepackage{amssymb}
\usepackage{graphicx}
\usepackage{graphics}
\usepackage{amsmath}
\usepackage{amsthm}

\def\L{L}
\def\l{l}
\def\T{O}
\def\S{O^\prime}

\def\X{X}

\def\Z{Z}

\newcommand{\bea}{\begin{eqnarray}}
\newcommand{\eea}{\end{eqnarray}}
\def\bi{\begin{itemize}}
\def\ei{\end{itemize}}
\def\bc{\begin{center}}
\def\ec{\end{center}}

\def\C{\hbox{$\mit I$\kern-.7em$\mit C$}}
\def\R{\hbox{$\mit I$\kern-.6em$\mit R$}}

\def\ket#1{|#1\rangle}
\newcommand{\one}{\mbox{$1 \hspace{-1.0mm}  {\bf l}$}}
\def\tr{\mathrm{tr}}
\def\ket#1{\left| #1\right>}
\def\bra#1{\left< #1\right|}

\newcommand{\proj}[1]{\ket{#1}\bra{#1}}

\def\RR{\mathbb{R}}
\newcommand{\poly}{{\rm poly}}

\begin{document}

\title{Compressed quantum simulation of the Ising model}
\author{B. Kraus}
\affiliation{Institute for Theoretical Physics, University of
Innsbruck, Austria}

\begin{abstract}

In [R. Jozsa, B. Kraus, A. Miyake, J. Watrous,
Proc. R. Soc. A {\bf 466}, 809-830 (2010)] it has been shown that a  match gate circuit running on $n$ qubits can be compressed to a universal quantum computation on $\log(n)+3$ qubits. Here, we show how this compression can be employed to simulate the Ising interaction of a $1$D--chain consisting out of $n$ qubits using a universal quantum computer running on $\log(n)$ qubits. We demonstrate how the adiabatic evolution can be realized on this exponentially smaller system and how the magnetization, which shows a quantum phase transition, can be measured.
\end{abstract}
\maketitle

The simulation of certain quantum system on a classical computer seems to be an unfeasible task. The reason for this is the exponential growth of the required resources, like space and time, as a function of the number of considered quantum systems. As conjectured by Feynman \cite{Fey82} and proven by LLoyd \cite{Llo96}, however, a quantum system can be used to simulate the behavior of another. The former one being such that the interactions between the systems are well--controllable and that the measurements can be performed sufficiently well. The suitability for the realization of such a quantum simulator has been shown for experimental schemes based on optical lattices or ion--traps \cite{JaViDu03,JaZo,PoCi04}. Recently, several experiments using for instance trapped ions \cite{FrScNatPhys08,KiCh10,IsMo11,Ibk2}, neutral atoms \cite{BlZw08} or NMR \cite{NMR,NMR2} have realized quantum simulations.

An important application of the quantum simulator is the study of the ground state properties of certain condensed matter systems. Quantum spin models are well suited for the investigations of quantum phase transitions \cite{Sa99}, which occur at zero temperature due to the change of some parameter, like the strength of the magnetic field, or pressure. The $1$D quantum Ising model, for instance, exhibits such a phase transition. It can be detected by measuring the magnetization as a function of the ratio between the interaction strength and the strength of the external magnetic field, which will be denoted by $J$ here. Since the Ising interaction is relatively simple, this is a good model for the experimental demonstration of quantum simulation, even though it can be simulated classically efficiently. 

One way to observe this quantum phase transition is to employ adiabatic evolution. The system, consisting out of $n$ qubits, is initially prepared in the (simple) ground state of the Hamiltonian corresponding to $J=0$. The adiabatic theorem tells us that if the parameter $J$ is slowly increased the system will be at any time in the ground state of the Hamiltonian for this value of $J$ \cite{AdTh}. Another way to transform the input state to the ground state of the quantum Ising model for the desired value of $J$ is to apply a specific gate, $U(J)$. In order to implement this gate, one usually approximates it using the Suzuki Trotter approximation \cite{Llo96}. Experimentally, the ground state properties, like the magnetization, $M(J)$, of the Ising model for $n=2$ \cite{FrScNatPhys08}, $n=3$ \cite{KiCh10} and $n=9$ \cite{IsMo11} 
have been recently observed. 

Here we use a different approach, which makes use of the fact that certain quantum circuits, the so--called matchgate (MG) circuits (MGC),
can be simulated by an exponentially smaller quantum system \cite{JoKr09}. We extend here this result and introduce new techniques to show that the evolution of the $1$D Ising model (including the measurement of the magnetization) of a spin chain consisting of $n$ qubits can be simulated by a compressed algorithm running only on $\hat{m}\equiv \log(n)$ qubits. Here and in the following $n$ is assumed to be a power of $2$ and the logarithm is taken in base $2$. More precisely, it is shown that the magnetization, $M(J)$, can be measured using the following algorithm: 1) First prepare the initial $\hat{m}$--qubit state $\rho_{in}=\one\otimes \proj{y_1}_{\hat{m}}$, with $\ket{y_1}=1/\sqrt{2}(\ket{0}-i\ket{1})$; 2) evolve the system up to a certain value of $J$ according to a specific unitary operator  $W(J)$ (see Eq. (\ref{EqW}) below); 3) measure the $\hat{m}$--th qubit in $y$--direction, i.e. $Y_{\hat{m}}$. The expectation value of $Y_{\hat{m}}$ coincides with the magnetization, $M(J)$ (for $n$ qubits) up to a factor $-1$.
The size of this circuit, i.e. the total number of single and two--qubit gates which are required to implement $W(J)$, coincides with the one required to implement $U(J)$ for the original circuit. Moreover, the error due to the Suzuki Trotter approximation is the same as the one of the original system, since we are simulating the gates exactly. Due to the fact that this compressed quantum computation corresponds to the simulation of the Ising model not only the magnetization, but also other quantities, like correlations can be measured.

This result allows for the experimental measurement of the quantum phase transition of very large systems with current technology. Consider for instance, experiments with ion-traps or NMR quantum computing where say $8$ qubits can be well--controlled \cite{Ibk1,IsMo11,NMR}. According to the results presented here such a system can be employed to simulate the interaction of $2^8=256$ qubits. Of course, for such a large system the phase transition can be well observed.

The outline of this paper is the following. First, we briefly recall the notion of MGs and some results related to MGCs. Then, we review some basic properties of the Ising model and its adiabatic simulation. After that we derive the MGs which are required for the adiabatic simulation and show how the symmetry of the Ising interaction can be used to compress the whole simulation into $\log(n)$ qubits.

Throughout the paper we use the following notation. $X,Y,Z$ denote the Pauli operators. If not stated differently, the subscript of an operator will always denote the system it is acting on, or the system it is describing. Normalization factors as well as the tensor product symbol will be omitted whenever it does not cause any confusion and $\one$ will denote the identity operator. 

Let us begin by reviewing the notion of MGs. A MG is a two--qubit gate which is of the form $ A\oplus B$, where $A$ is acting on the even parity subspace, i.e. on span $\{\ket{00},\ket{11}\}$ and $B$ is acting on the odd parity subspace, i.e. on span $\{\ket{01},\ket{10}\}$. Both, $A$ and $B$ are unitary and have the same determinant. A MGC is a quantum circuit which fulfills the following conditions: (i) the MGs act only on nearest neighbor (n.n.) qubits; (ii) the input state is any computational basis state (iii) the output is a final measurement in the computational basis on any single qubit. In \cite{valclsim} (see also \cite{jm08,terdiv}) it has been shown that the output of any MGC can be classically efficiently simulated. In the following ``n.n. MG'' will be simply called ``MG''.

In order to review how the efficient classical
simulation is actually achieved we begin by introducing the $2n$ hermitian operators on $n$-qubits \cite{JordanWigner}:
\begin{equation} \begin{array}{cccc}
c_1=X\,I\cdots I  & \quad\cdots\quad &
c_{2k-1}= Z\cdots Z\,X\,I\cdots I & \quad\cdots
\\
c_2=Y\,I\cdots I &  \quad\cdots\quad &
\,\,c_{2k}\,\,\,\, = Z\cdots Z\,Y\,I\cdots I & \quad\cdots
\end{array} \nonumber.
\end{equation} Here, $X$ and $Y$ are in the $k$-th slot for
$c_{2k-1}$ and $c_{2k}$, and $k$ ranges from 1 to $n$.
It can be easily verified that these operators satisfy the
anti-commutation relations $\{ c_j, c_l \} \equiv c_j c_l + c_l c_j = 2
\delta_{j,l} I$ for $j,l = 1, \ldots ,2n.$

Let $U$ be a MG acting on the qubits $k$ and $k+1$. Then
\begin{equation}\label{mgrot}
  U^\dagger c_j U = \sum_{l=1}^{2n} R_{jl} c_l,
\end{equation}
where $R\in SO(2n,\RR )$ is a special orthogonal matrix with $R_{jl}=\delta_{j,l}$ $\forall l\not\in \{2k-1,2k,2k+1,2k+2\}$ \cite{jm08}. Consider now a MGC $U=U_N\ldots U_1$ acting on the input state $\ket{\Psi_{\rm in}}$. Let us denote by $R_i$ the real orthogonal matrices associated to $U_i$ (see Eq. (\ref{mgrot})) and by $R$ the orthogonal matrix associated to $U$, i.e. $R=R_N\ldots R_1$. Noting that $Z_k=-i c_{2k-1} c_{2k}$ we obtain for the final $Z$ measurement on qubit $k$
\bea \label{p01a}
\langle Z_k\rangle
  = \bra{\Psi_{\rm in}} U^\dagger Z_k U\ket{\Psi_{\rm in}}=\bra{2k-1}RSR^T\ket{2k},\eea
where $S_{jl}=\bra{\Psi_{\rm in}} -i c_jc_l\ket{\Psi_{\rm in}}$ for $j\neq l$ and $S_{jj}=0$ \footnote{We can choose $S_{jj}=0$ wlog because of the orthogonality of $R$.}. For instance, for $\ket{\Psi_{\rm in}}=\ket{0}^{\otimes n}$ the corresponding $S$ matrix would be $S=i\oplus_k Y_k=\one \otimes iY$, where $\one$ denotes the (unnormalized) identity matrix on $n$ dimensions. For any computational basis state $S$ can be computed efficiently. Hence, the simulation of the MGC, which consists then only of multiplications of $2n\times 2n$ matrices, which can be efficiently determined, can be performed classically efficiently \footnote{Let us note here that the n. n. restriction on MGs is necessary for the classical simulability of the circuits. In fact, n.n. and next-n.n. MGs are already universal for quantum computation \cite{jm08}.}.

In \cite{JoKr09} we showed that a MGC on $n$ qubits can be compressed to a universal quantum computation running on $\log(n)+3$ qubits. If $N$ denotes the size of the MGC, i.e. the number of single and two--qubit gates, then the size of the compressed algorithm is ${\cal O}(N\log(n))$. Thus, any polynomial-sized MGC, can be
simulated by a universal quantum computer of $\poly(n)$ size and
exponentially compressed width $O(\log n)$.
The idea was to apply the controlled gate, $\Lambda U=\proj{0}\otimes \one+\proj{1}\otimes U$, with $U=S^{-1}RSR^T$, to the input state $\ket{+}\ket{0}^{\otimes \log(n)}$. Measuring then the first system in $X$--direction leads to the desired result, $\bra{1}RSR^{T}\ket{2}$ \footnote{We used here the fact that
any MGC with input state $\ket{x_1,\ldots x_n}$, with $x_i\in \{0,1\}$ and $Z$--measurement on qubit $k$ can be mapped to an equivalent MGC with input state $\ket{0,\ldots 0}$ and $Z$--measurement on qubit $1$ \cite{JoKr09}.}. Due to the fact that the required classical side--computation can be performed on $\log$--space, the computation is indeed performed by the exponentially smaller quantum computer.

Next, some basic properties of the Ising model are reviewed. We consider the Hamiltonian, $H(J)=H_0+J H_1$, to describe the $1$D Ising model with open boundary conditions.
Here, $H_0=\sum_{i=1}^n \Z_i$ and $H_1=\sum_{i=1}^{n-1} \X_i\otimes X_{i+1}$. In Appendix A the spectrum of the Hamiltonian and the corresponding non--degenerate ground state are re--derived \cite{JordanWigner,Sa99}. Since we want to compress the computation we have to use here open boundary conditions, since, e.g. periodic boundary conditions would not correspond to n.n. MGs.

The quantum Ising model shows a quantum phase transition at the critical value $J=1$. At this point the second derivative of the magnetization is no longer continuous. One way to experimentally observe this phase transition is to use adiabatic evolution \cite{AdTh} \footnote{The speed with which the parameter $J$ can be varied is upper bounded by a function proportional to $1/\Delta^2$, where $\Delta=E_1-E_0$ denotes the energy gap between the ground state and the first excited state. Note that even if $\Delta$ is unknown, experimentally it can be easily checked if the evolution was indeed slow enough \cite{MuCi03}.}. To describe the evolution we discretize the Hamiltonian $H(J)$ into $\L+1$ steps, as $H(\l)$, where $\l$ goes from $0$ to $\L$ and $H(0)=H_0$ and $H(\L)=H_0+J_{max}H_1$. In the case of the Ising Hamiltonian we have $H(\l)=H_0+J(\l)H_1$, with $J(\l)=J_{max}\l/\L $. To obtain the ground state of the Hamiltonian $H(J)$, the system has to evolve according to the unitary operator $\tilde{U}(J)= \prod_{\l=1}^{L(J)} \tilde{U}_\l,$ where $L(J)=JL/J_{max}$ and $\tilde{U}_\l=e^{-iH(\l)\Delta t}$. Here, $\Delta t=T/(\L+1)$, where $T$ is the time of the evolution. The adiabatic limit is achieved for $T,\L\rightarrow \infty$ ($\Delta t\rightarrow 0$).

In a next step we use the Suzuki Trotter expansion to approximate $\tilde{U}_\l$ up to second order in $\Delta t$, i.e. $\tilde{U}_\l\approx e^{-iH_0\Delta t/2}e^{-iJ(l) \Delta t H_1\Delta t}e^{-iH_0\Delta t/2}+{\cal O}(\Delta t^2)$. Thus, we approximate the unitary $\tilde{U}(J)$ by \bea \label{UJ} U(J)= \sqrt{V_0} \prod_{\l=1}^{L(J)} U_\l \sqrt{V_0}^\dagger,\eea
where $V_0=e^{-iH_0\Delta t}$ and $U_\l=V_\l V_0$ with $V_\l=e^{-iJ(\l) H_1\Delta t}$.

Hence, the procedure to simulate the adiabatic evolution and to observe the quantum phase transition is the following. First, the input state $\ket{0}^{\otimes n}$, which is the ground state of $H_0$, is prepared. Then the system is evolved according to the unitary $U(J)$. After that, the magnetization as a function of $J$, $M(J)=1/n\sum \bra{\Psi_J} Z_k\ket{\Psi_J} $, where $\ket{\Psi_J}=U(J)\ket{0}^{\otimes n}$ denotes the ground state of the Hamiltonian $H(J)$, is measured.

We will compress now the circuit explained above, which is running on $n$ qubits to an equally efficient one which is running only on $\log(n)$ qubits. Note that for the standard classical simulation of a MGC, $2n$ dimensions are required, which corresponds to the Hilbert space of  $m=\log(2n)=\log(n)+1$ qubits (see Eq. (\ref{p01a})). However, we will see that one can compress the algorithm even to $\log(n)$ qubits due to the symmetry of the Ising model. It should be noted here, that the compressed algorithm does indeed simulate the adiabatic evolution of the system. That is, it is not only possible to measure the magnetization using the compressed quantum computation, but also for instance correlation functions, like $X_i\otimes X_{i+1}$, could be measured. In this case, only the measurement would change, but the evolution would be the same.

First we derive the rotations $R_i$, which correspond to the MGs $V_0\equiv e^{-iH_0\Delta t}$ and $V_\l\equiv e^{-iJ(\l) H_1\Delta t}$ (see Eq. (\ref{mgrot})). For a MG, $U=e^{-iH t}$ we have $H=i\sum_{j,k=1}^{2n} h_{jk} c_jc_k$, where $h$ is antisymmetric. The corresponding rotation is then of the form $R=e^{4h t}$ \cite{jm08}. It is easy to verify that $h_0=-1/2\sum_{k=1}^n(\ket{2k-1}\bra{2k}-\ket{2k}\bra{2k-1})$ and $h_1=-1/2\sum_{k=1}^{n-1}(\ket{2k}\bra{2k+1}-\ket{2k+1}\bra{2k})$. In order to make the connection to qubit systems we use the binary notation, $k=\sum_{l=1}^{m} 2^{m-l} k_l +1$ and write $\ket{k}=\ket{k_1,\ldots, k_m}$. Then we have $h_0=i/2\one \otimes Y_m$ and therefore $R_0\equiv e^{4h_0 \Delta t}=\one \otimes e^{i 2 \Delta t Y_m}$ and $R_l\equiv e^{4J(l)h_1 \Delta t}= (1-\cos(2J(l) \Delta t))(\proj{1}+\proj{2n})+\cos(2 J(l) \Delta t )\one+\sin(2J(l) \Delta t)\sum_{k=1}^{n-1}\ket{2k+1}\bra{2k}-\ket{2k}\bra{2k+1},$ with $J(l)= J_{max}l/L$. Thus, the compressed evolution which corresponds to the adiabatic evolution, $U(J)$, is given by $R(J)=\sqrt{R_0}\prod_{\l=1}^{\L (J)} T_\l\sqrt{R_0}^{-1}$, with $T_\l=R_\l R_0$ and $\L(J)=JL/ J_{max}$.  Since the ground state of the Hamiltonian $H(0)$, i.e. the initial state, is $\ket{0}^{\otimes n}$ the corresponding matrix $S$ (see Eq. (\ref{p01a})), is given by $S=i\one\otimes Y_m$. Putting everything together we have $\langle Z_k\rangle=\tr( U(J)^\dagger Z_k U(J) \proj{0^{\otimes n}})=\bra{2k}R(J)SR(J)^T\ket{2k-1}$, where $R(J)$ and $S$ are defined above. Note that $U(J)$ is acting on $n$ qubits, whereas $R(J)$ is only acting on $m=\log(n)+1$ qubits. In the following we will use a superscript to indicate the dimension of the Hilbert space an operator is acting on. For instance the operators $R^m_0, R^m_1$ and $S^m$ denote the operators $R_0,R_\l$ acting on ${\cal H}=(\C^{2})^{\otimes m}$, i.e. on the Hilbert space of $m$ qubits.

In order to determine the magnetization we can either measure all expectation values $\langle Z_k\rangle$ and average over them, or, we use the fact that for large values of $n$ the boundary conditions will not play an important role and that all expectation values will be approximately equal to $Z_{k\approx n/2}$, like in the case of periodic boundary conditions. In this case, we would simply measure only one expectation value. Here, we will compute the average since we are also dealing with systems composed out of a few qubits only. To this end we use the fact that for any real antisymmetric matrix $A$ we have $\sum_{k=1}^n \bra{2k-1}A\ket{2k}=1/2\tr(A \one \otimes iY_m)$ \footnote{ This can be easily seen by using that $\ket{2k-1}=\ket{k_1,\ldots k_{m-1},0}$ and $\ket{2k}=\ket{k_1,\ldots k_{m-1},1}$, and therefore $\one \otimes iY=\sum_{k_1,\ldots k_{m-1}}\proj{k_1,\ldots k_{m-1}}\otimes (\ket{1}\bra{0}-\ket{0}\bra{1})=\sum_{k=0}^n \ket{2k}\bra{2k-1}-\ket{2k-1}\bra{2k}$ and that $\bra{2k-1}A\ket{2k}=-\bra{2k}A\ket{2k-1}$.}. Since $RS R^T$  has the properties of matrix $A$ we have $M(J)=
i/(2n)\tr(R(J) S R(J)^T \one \otimes Y_m)$. Using now that $R_0$ commutes with $S=i\one\otimes Y_m$ and the orthogonality of the matrices we find
\bea \label{MJm1} M(J)=-1/(n) \tr(\hat{R}(J) \rho_{in} \hat{R}(J)^T \one \otimes Y_m),\eea

with 
\bea \hat{R}(J)=\prod_{\l=1}^{\L(J)} T_\l\equiv \prod_{\l=1}^{\L(J)} R_\l R_0\eea
 and $\rho_{in}=1/2(S/i+\one)=\one \otimes \proj{y_1}$, with $\ket{y_1}$ being the eigenstate to eigenvalue $1$ of $Y$.

In contrast to \cite{JoKr09} we act here only on $m$, not $m+1$ qubits. The reason for this is that we do not apply the controlled unitary $\Lambda S^{-1}\hat{R}(J)S \hat{R}(J)^T$, but we simply let the system evolve according to $\hat{R}(J)$. Note that this also allows for a simplification in the evolution and that this idea can be used for the simulation of any MGC. Due to Eq. (\ref{MJm1}), $M(J)$ can be measured by preparing the $m$--qubit input state $\rho_{in}=\one \otimes \proj{y_1}$ then applying $\hat{R}(J)$ for the desired value to $J$ and measuring $Y_m$.

In Appendix B we show that due to the symmetry of the Ising model 
the algorithm can be even further compressed to $\hat{m}=m-1=\log(n)$ qubits. More precisely, it is shown that \bea \label{Magfinal} M(J)=-\frac{2}{n}\tr(W(J) \rho_{in}^{\hat{m}}W(J)^\dagger \one\otimes Y_{\hat{m}}),\eea
where $\rho_{in}^{\hat{m}}=\one_{m-2}\otimes \proj{y_1}_{\hat{m}}$ and the $\hat{m}$ qubit operators \bea \label{EqW} W(J)=\prod_{\l=1}^{\L(J)} \T_\l (R_0^{\hat{m}})^T \eea with
$R_0^{\hat{m}}=\one\otimes e^{i 2 \Delta t Y_{\hat{m}}}$, and $
\T_\l=U_d (R_\l^{\hat{m}})^T$. Here, $U_d$ denotes the $\hat{m}$--qubit phase gate which leaves all computational basis states but $\ket{1}^{\otimes \hat{m}}$ unchanged and the state $\ket{1}^{\otimes \hat{m}}$ is mapped to $e^{i J 2 \Delta T}\ket{1}^{\otimes \hat{m}}$.

Thus, the whole quantum simulation can be performed on $\log(n)$ qubits. Explicitly, the compressed algorithm for the simulation of the Ising model reads: 1) prepare the initial $\hat{m}$--qubit state $\rho_{in}=\one\otimes \proj{y_1}_{\hat{m}}$. 2) evolve the system up to a certain value of $J$ according to the unitary operator  $W(J)$ [Eq. (\ref{EqW})]. 3) measure $Y_{\hat{m}}$.
The expectation value of $Y_{\hat{m}}$ equals $-M(J)$ (for $n$ qubits).
In Fig. $1$ (lower insert) the magnetization as a function of $J$ for different values of the system size is shown. The phase transition can be clearly seen for values of $\hat{m}\geq 5$, i.e. for $n\geq 32$. To see how many steps in the Trotter expansion are required we compared in Fig. $1$ (Fig. $2$) the exact value of $M(J)$ with the one obtained by the simulated evolution for $\hat{m}=3$ ($\hat{m}=8$). There, $M(J)$ for different values of the total evolution time $T$, keeping $\Delta t=0.05$ fixed, is shown. Considering the evolution for $\hat{m}=3$ (Fig. $1$) it can be seen that choosing $T=30$, which amounts to $600$ steps in the Trotter expansion, suffices for the approximation. Recall that the same number of steps would be required for the simulation using the exponentially larger system. However, the effect of errors will be much smaller in the compressed simulation. In order to demonstrate that, we consider in Fig $1$ (upper insert) the error $(e^{i\alpha Z})^{\otimes \log(n)}$ being applied after each step, $O_l R_0^T$ in the compressed algorithm. The errors, $\alpha$, are chosen randomly between $0$ and $x\equiv 10^{-2}$ (green line) and $0$ and $x=10^{-3}$ (red line). It should be noted that an error $x\leq 10^{-3}$ basically does not affect the result.

\begin{figure}[h!]
\begin{center}
  \includegraphics[height=0.25\textheight]{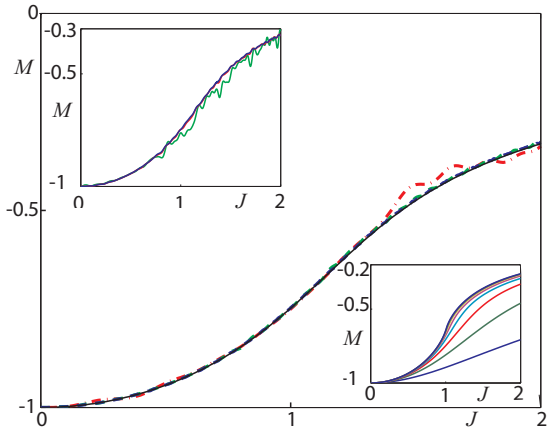}
  \caption{The magnetization as a function of $J$ for $\hat{m}=3$ (i.e. $n=8$). The solid line (online black) shows the exact value. The dashed--dotted line (red) shows the magnetization obtained for the total evolution time $T=10$ and the number of Trotter steps, $L=200$, ($\Delta t=0.05$); the dashed line  (green) for $T=30$ and $L=600$ ($\Delta t=0.05$). The magnetization for $T=100$ and $L=2000$ ($\Delta t=0.05$) basically coincides with the exact value. The upper insert shows $M(J)$ for $T=30, \Delta t=0.05$ ($L=600$) solid line (blue) for different values of the error (see main text): lower line (green) for $x=10^{-2}$ and dashed--dotted line (red) for $x=10^{-3}$. The lower insert shows the exact value of $M(J)$ for $\hat{m}=1$ (lowest line) to $\hat{m}=6$ and for $\hat{m}=8$ (most upper line). }\label{figmagN8}
\end{center}
\end{figure}

\begin{figure}[h!]
\begin{center}
  \includegraphics[height=0.25\textheight]{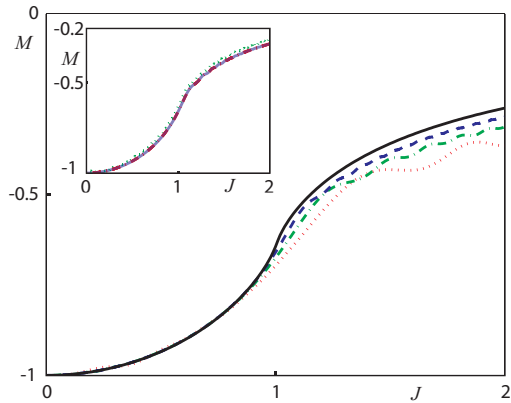}
  \caption{The same plot as in Fig. (\ref{figmagN8}) but for $\hat{m}=8$, i.e. $N=2^8$. The insert is plotted for $T=100$, and $L=2000$ ($\Delta t=0.05$), where the adiabatic evolution approximates the exact solution well.}\label{figmagN256}
\end{center}
\end{figure}

In summary, we have shown that the simulation of the Ising model of $n$ qubits can be performed on a system consisting only out of $\log(n)$ qubits. The compressed quantum algorithm is an exact simulation of the adiabatic evolution: Whereas on the $n$--qubit system the unitary operation $U(J)$ (see Eq. (\ref{UJ})) has to be applied for a certain value of $J$ before the magnetization is measured, the unitary operation $W(J)$ (see Eq. (\ref{EqW})) has to be applied to the exponentially smaller system, before $Y_{\hat{m}}$ is measured. The error due to the Suzuki Trotter approximation is the same as for the original system, since we are simulating the gates exactly. However, in experiments, the error which will occur by implementing a gate on $\log(n)$ qubits instead of $n$ qubits will affect the result less (see upper insert of Fig. $1$ and Fig. $2$).
Furthermore, the size of the circuit does not increase compared to the original adiabatic simulation. The (non--compressed) adiabatic evolution of the Ising model can be decomposed into the following MGs. First apply $e^{i \Delta t Z_i}$ to each system $i$ then the commuting two--qubit gates $e^{iJ(l) \Delta t X_i\otimes X_{i+1}}$ for $i$ odd and afterwards the same gates for $i$ even are applied. Thus, the number of required gates here is ${\cal O}(n)$. Since $W(J)$, is just a $\log(n)$--qubit gate, it requires at most ${\cal O}(n)$ basic operations. For instance the implementation of $R_0^T=\one\otimes e^{-i2\Delta tY_m}$, which corresponds to the $n$ gates $e^{i \Delta t Z_i}$, requires only one gate. It is also interesting to compare the circuits of size $m$ and $m-1$. One only needs to change $R_0^m$ to $(R_0^{m-1})^T$ and $R_l^m$ to $U_d(R_l^{m-1})^T$. It is due to the phase gate, $U_d$, why the circuit cannot be further compressed. Finally, note that this compression of quantum simulation is not restricted to the Ising model, but can also be used for instance to simulate the $XY$--model, as already suggested in \cite{JoKr09}. The results presented here can also be used to study and experimentally simulate quench--induced quantum phase transitions, which would lead to new and interesting observations \cite{ZuZo05}.

I would like to thank J. I. Cirac for helpful discussions and acknowledge support by the FWF (START Preis) and SFB-FOQUS.


\section*{Appendix A: Exact diagonalization of $H(J)$}

Here, we briefly review a method to compute the eigenvalues and the magnetization of the Ising model \cite{Sa99}. We introduce the fermionic operators $a_k^\dagger=Z...Z\sigma_k^+ \one=1/2(c_{2k-1}-ic_{2k})$ and $a_k=Z...Z\sigma_k^- \one=1/2(c_{2k-1}+ic_{2k})$. It is easy to see that those operators fulfill the anticommutation relations $\{a_i,a_j\}=0$ and $\{a_i,a_j^\dagger\}=\delta_{ij}\one$. Since $\sigma_l^+=(-1)^{\sum_{k=1}^{l-1}a_k^\dagger a_k}a_l^\dagger$ we have $X_l=(-1)^{\sum_{k=1}^{l-1}a_k^\dagger a_k}(a_l^\dagger+a_l)$ and $Z_l=\one-2a_l^\dagger a_l$. Thus, $H=\vec{a}^\dagger M \vec{a},$ where $\vec{a}=(a_1,a_1^\dagger,\ldots a_n,a_n^\dagger)$ and \bea M=\left(\begin{array}{ccccccc}A&B&0&\ldots&\ldots&\ldots&0\\ \nonumber
B^T&A&B&0&\ldots&\ldots&0 \\ \nonumber0&B^T&A&B&0&\ldots& 0\\ \nonumber ...&&&&&&\\ \nonumber 0&\ldots &\ldots&0&B^T&A&B \\ \nonumber 0&\ldots&\ldots&\ldots& 0&B^T&A\end{array}\right),\eea
with $A=-Z$ and $B=-J/2(Z+iY)$. Since $(\one_n\otimes X) M (\one_n\otimes X)=-M$ the eigenvalues of $M$ always occur in pairs, like $\lambda_i,-\lambda_i$. Diagonalizing this symmetric and real matrix $M=UDU^T$, where $D=\mbox{diag}(d_1,\ldots d_{2n})$, with $d_1=\lambda_n, d_2=-\lambda_n, d_3=\lambda_{n-1}, d_4=-\lambda_{n-1},\ldots d_{2n-1}=\lambda_0, d_{2n}=-\lambda_0$, where $\lambda_n\geq \lambda_{n-1}\geq \ldots \geq \lambda_0\geq 0$ and $U$ an orthogonal matrix, which depends on $J$ , we find $H=\vec{b}^\dagger D\vec{b}$, with $\vec{b}=U^\dagger \vec{a}$. Hence, $H(J)$ can be expressed in the new fermionic operators, $b_i$ as $H(J)=\sum_{l=1}^{n} d_{2l}\one+(d_{2l-1}-d_{2l})b_{2l-1}^\dagger b_{2l-1}$ with the ground state energy $\sum_{l=1}^{n} d_{2l}$
and the corresponding (non--degenerate) eigenstate, the vacuum states of the modes $b_i$, i.e. $\ket{\Psi(J)}=\ket{0\ldots 0}_{b_i(J)}$. It is important to note here that the ground states energy is not degenerate. 

Let us now compute the magnetization $M(J)=1/n\bra{\Psi(J)}\sum_l Z_l \ket{\Psi(J)}$. Since $Z_l=-2a_l^\dagger a_l+\one$ and $a_l=(\vec{a})_{2l-1}$ and $a^\dagger_l=(\vec{a})_{2l}$ we have $\sum_l a_l^\dagger a_l=\vec{a}^\dagger R\vec{a}=\vec{b}^\dagger U^\dagger R U\vec{b}$, where $R=\sum_l\proj{2l-1}$. Thus, $\bra{\Psi(J)}\sum_l a_l^\dagger a_l \ket{\Psi(J)}=\bra{\Psi(J)}\vec{b}^\dagger U^\dagger R U\vec{b}\ket{\Psi(J)}$, and therefore $M(J)=-2/n\sum_l (U^\dagger R U )_{2l,2l}+1$, which shows a quantum phase transition at the critical point $J=1$, as there the second derivative of $M(J)$ is not continuous. 

\section*{Appendix B: Compressed circuit of width $\log(n)$}

In order to prove that the Ising model can be simulated by a quantum circuit of width $\log(n)$ we introduce the following unitary operator: \bea V=1/\sqrt{2} (\proj{0}_1\otimes \one \otimes Z_{m} -i\proj{1}\otimes \one \\ \nonumber -i\ket{0}\bra{1} \otimes X^{\otimes m-1}+i\ket{1}\bra{0} \otimes X^{\otimes m-2}\otimes Y_{m}.\eea This unitary can also be written as $V=1/\sqrt{2}\sum_{k=1}^{2n} \alpha_k \proj{k}+\beta_k\ket{2n-k+1}\bra{k},$ where $\alpha_k=(-1)^{k+1}$ and $\beta_k=(-1)^k $ $\forall k\leq n$ and $\alpha_k=\beta_k=-i$ $\forall k>n$. Below we show that $\tilde{R}^m_0\equiv VR^m_0 V^\dagger=e^{-i 2 \Delta t Z_1\otimes \one \otimes Y_{m}} = \proj{0}_1 \otimes (R_0^{\hat{m}})^T+\proj{1}_1 \otimes R_0^{\hat{m}}$.
Furthermore, we have $\tilde{R}_\l\equiv VR_\l V^\dagger= \proj{0}_1 \otimes \T_\l + \proj{1}_1 \otimes \S_\l ,$  where  $\T_\l=U_d (R_\l^{\hat{m}})^T$ with $U_d=\one+(e^{i2J\Delta t}-1)\proj{1}^{\hat{m}}$ and $\S_\l=X^{\otimes \hat{m}} (\T_\l)^\ast X^{\otimes \hat{m}}$. Note that both, $\tilde{R}_0$ and $\tilde{R}_\l$ are block diagonal, which implies that also $\tilde{R}(J)=V\hat{R}(J)V^\dagger=\prod_{\l=1}^{\L(J)} (\tilde{R}_\l \tilde{R}_0)$ is block--diagonal. More precisely, we have $\tilde{R}(J)=\proj{0}_1 \otimes \prod_\l [\T_l (R_0^{\hat{m}})^T] +\proj{1}_1 \otimes \prod_\l [\S_\l R_0^{\hat{m}}]\equiv \proj{0}_1 \otimes W_0 +\proj{1}_1 \otimes W_1$. It is straightforward to show that $W_1=X^{\otimes \hat{m}} W_0^\ast X^{\otimes \hat{m}}$. Using that $VS^mV^\dagger=-Z_1\otimes S^{\hat{m}}$, we obtain for the magnetization $M(J)=-1/(2n)\tr((\proj{0}_1 \otimes W_0 Y_m W^\dagger_0+\proj{1}_1 \otimes W_1 Y_m  W_1^\dagger )Y_m))=-1/(n)\tr((W_0 \one\otimes Y_{\hat{m}} W^\dagger_0 \one\otimes Y_{\hat{m}})$. For the last equality we used the fact that $\tr(W_0 \one\otimes Y_{\hat{m}} W^\dagger_0 \one\otimes Y_{\hat{m}})=\tr(W_1 \one\otimes Y_{\hat{m}} W^\dagger_1\one\otimes Y_{\hat{m}})$, which can be easily verified by noting that $\tr(W_0 \one\otimes Y_{\hat{m}} W^\dagger_0 \one\otimes Y_{\hat{m}})$ is real \cite{footnote}.
This proves the validity of Eq. (\ref{Magfinal}).

The fact that $\tilde{R}_0$ and $\tilde{R}_1$ are block--diagonal can be easily seen as follows. We write $R_0=\cos(2\Delta t)\one+ \\ \nonumber \sin(2\Delta t)\sum_{k=1}^{n/2} \ket{2k}\bra{2k-1}-\ket{2n-2k+1}\bra{2n-2k+2}- h.c.$. Using that $\ket{2n-k+1}=\X^{\otimes m}\ket{k}$ it is easy to show that $ \tilde{R}_0=VR_0V^\dagger=\cos(2\Delta t)\one- \sin(2\Delta t)\sum_{k=1}^{n/2} \ket{2k}\bra{2k-1}-\ket{2n-2k+1}\bra{2n-2k+2}- h.c.=e^{-i2\Delta t Z_1\otimes \one \otimes Y_m}$. Similarly, we write $R_l= \cos(x)\one+ (1-\cos(x))(\proj{1}+\proj{2n})+\sin(x)\sum_{k=1}^{n/2-1}(\ket{2k+1}\bra{2k}-\ket{2n-2k}\bra{2n-2k+1}-h.c.)+ \sin(x)(\ket{n+1}\bra{n}-\ket{n}\bra{n+1})$, where $x=2 J(l) \Delta t $. Then we find $\tilde{R}_l=\cos(x)\one+ (1-\cos(x))\proj{1}+\sin(x)\sum_{k=1}^{n/2-1}(\ket{2k}\bra{2k+1}-\ket{2k+1}\bra{2k}
+i \sin(x)\proj{n}+ X^{\otimes m} \sin(x)\sum_{k=1}^{n/2-1}(\ket{2k}\bra{2k+1}-\ket{2k+1}\bra{2k} X^{\otimes m}$. For any $k$ such that $k\leq n/2-1$ $\ket{k}=\ket{0,k_2,\ldots, k_m}$ and therefore $\tilde{R}_l=\proj{0}\otimes \T_l+\proj{1}\otimes \S_l$, where $\T_l=\cos(x )\one+ (1-\cos(x))\proj{1}+\sin(x)\sum_{k=1}^{n/2-1}(\ket{2k}\bra{2k+1}-\ket{2k+1}\bra{2k}
+i \sin(x)\proj{n}$ and $\S=X^{\otimes m}(\T)^\ast X^{\otimes m}$. Note that $\T_\l=U_d (R_\l^{\hat{m}})^T$, where $U_d$ is the $\hat{m}$--qubit phase gate, $\one+(e^{x}-1)\proj{1}^{\otimes \hat{m}}$. Thus, both, $\tilde{R}_0$ and $\tilde{R}_l$ are block--diagonal. 

\begin{thebibliography}{99}

\bibitem{Fey82} R. P. Feynman, Int. J. Theor. Phys, {\bf 21},467--488 (1982).

\bibitem{Llo96} S. Llody, Science, {\bf 273}, 1073 (1996).

\bibitem{JaZo} D. Jaksch and P. Zoller, Annals of Physics {\bf 315}, 52
(2005). 

\bibitem{PoCi04} D. Porras, J. I. Cirac, Phys. Lett. {\bf 92}, 207901 (2004).

\bibitem{JaViDu03} E. Jan$\acute{e}$, G. Vidal, W. D\"ur, P. Zoller, J.I. Cirac, Quant. Inf. Comp., {\bf 3}, 1, 15 (2003).


\bibitem{FrScNatPhys08} A. Friedenauer, H. Schmitz, J. Glueckert, D. Porras and T. Schaetz.
Nature Physics {\bf 4}, 757 (2008).


\bibitem{KiCh10} K. Kim, M.-S. Chang, S. Korenblit, R. Islam, E. E. Edwards, J. K. Freericks, G.-D. Lin, L.-M. Duan3, C. Monroe, Nature, {\bf 465}, 590 (2010).

\bibitem{IsMo11} R. Islam et. {\it al},  Nature Communications, {\bf 2}, 377 (2011).

\bibitem{Ibk2} B. P. Lanyon et. {\it al.}, Science, 1208001 (2011).

\bibitem{BlZw08} for a recent review see I. Bloch, J. Dalibard, W. Zwerger, Rev. Mod. Phys. {\bf 80}, 885 (2008).

\bibitem{NMR} J. Zhang, T.-Ch. Wei, R. Laflamme, Phys. Rev. Lett. {\bf 107}, 010501 (2011).

\bibitem{NMR2} J. Zhang, M.-H. Yung, R. Laflamme, A. Aspuru-Guzik, J. Baugh, quant-ph/11083270 (2011).

\bibitem{Sa99} S. Sachdev, {\it Quantum Phase Transitions}, Cambridge University Pree, Cambridge, 1999.

\bibitem{AdTh} see for instance A. Messiah, Quantum Mechanics, Vol. II, Amsterdam, North Holland, New York, Wiley (1976).

\bibitem{JoKr09} R. Jozsa, B. Kraus, A. Miyake, J. Watrous,
Proc. Roy. Soc. A {\bf 466}, 809 (2010).

\bibitem{Ibk1} H. H\"affner et {\it al.}, Nature, {\bf 438}, 643 (2005); T. Monz et al, Phys. Rev. Lett. 106 , 130506 (2011).

\bibitem{valclsim} L. Valiant, SIAM J. Computing, {\bf 31:4},
1229 (2002).

\bibitem{jm08} R. Jozsa, and A. Miyake, Proc. Roy. Soc.
(Lond) {\bf A464}, 3089 (2008).

\bibitem{terdiv} B. Terhal and D. DiVincenzo, Phys. Rev. A
{\bf 65}, 032325 (2002); E. Knill, quant-ph/0108033 (2001).

\bibitem{JordanWigner} P. Jordan and E. Wigner, Zeitschrift
f\"{u}r Physik, {\bf 47}, 631 (1928). 

\bibitem{MuCi03} V. Murg, J. I. Cirac, Phys. Rev. A {\bf 69}, 042320 (2004).

\bibitem{ZuZo05} W. H. Zurek, U. Dorner, P. Zoller, Phys. Rev. Lett {\bf 95}, 105701 (2005).

\bibitem{footnote} This follows form the fact that for any operator $A$ and any system $k$ the following holds $_k\bra{0} A Y_k A^\dagger Y_k \ket{0}_k=(_k\bra{1} A Y_k A^\dagger Y_k \ket{1}_k)^\ast$.

\end{thebibliography}
\end{document}